\documentclass[twocolumn,showpacs,preprintnumbers,pra]{revtex4}
\usepackage{amssymb}
\usepackage{amsmath}
\usepackage{graphicx}
\usepackage{dcolumn}
\usepackage{bm}

\begin{document}

\title{The roles of a quantum channel on a quantum state}
\author{Lin Wang }
\author{ Chang-shui Yu}
\email{quaninformation@sina.com;  ycs@dlut.edu.cn}

\affiliation{School of Physics and Optoelectronic Technology, Dalian University of
Technology, Dalian 116024, P. R. China}

\begin{abstract}
When a quantum state undergoes a quantum channel, the state will be
inevitably influenced. In general, the fidelity of the state is reduced, so
is the entanglement if the subsystems go through the channel. However, the
influence on the coherence of the state is quite different. Here we present
some state-independent quantities to describe to what degree the fidelity,
the entanglement and the coherence of the state are influenced. As
applications, we consider some quantum channels on a qubit and find that the
infidelity ability monotonically depends on the decay rate, but in usual the
decoherence ability is not the case and strongly depends on the channel.
\end{abstract}

\maketitle

\section{Introduction}

Quantum coherence or quantum superposition is one of the most fundamental
feature of quantum mechanics that distinguishes the quantum world from the
classical world. It is one of the main manifestation of quantumness in a
single quantum system. For a composite quantum system, due to its tensor
structure, quantum superposition could directly lead to quantum
entanglement, another intriguing feature of quantum mechanics and the very
important physical resource in quantum information processing [1]. In fact,
safely speaking, quantum coherence is one necessary condition for almost all
the mysterious features of a quantum state. For example, both entanglement
and quantum discord that has attracted much attention recently [1-10], have
been shown to be quantitatively related to some special quantum coherence
[11,12]. However, when a quantum system undergoes a noisy quantum channel or
equivalently interacts with its environment, the important quantum feature,
i.e., the quantum decoherence, could decrease. It is obvious that whether
decoherence happens strongly depends on the quantum channel and the state
itself, but definitely a quantum channel describes the fate of quantum
information that is transmitted with some loss of fidelity from the sender
to a receiver. In addition, if the subsystem of an entangled state passes
through such a channel, disentangling could happen and go even more quickly.

Decoherence as well as disentangling for composite systems has never been
lack of concern from the beginning. A lot of efforts have been paid to
decoherence in a wide range such as the attempts of understanding of
decoherence [13,14], the dynamical behaviors of decoherence subject to
different models [15-20], the reduction or even prevention of decoherence
[21-23], the disentangling through various channels and so on [24-26].
Actually, most of the jobs can be directly or indirectly summarized as the
research on to what degree a noisy quantum channel or the environment
influences (mostly destroys) the coherence, the fidelity or entanglement of
the given quantum system of interests. So it is important and interesting to
consider how to effectively evaluate the ability of a quantum channel that
leads to decoherence, the loss of fidelity of a quantum state, or
disentangling of a composite system, in particular, independent of the state.

In this paper, we address the above issues by introducing three particular
measure, the decoherence power, the infidelity power and the disentangling
factor to describe the abilities, respectively. This is done by considering
how much fidelity, coherence or entanglement (for composite systems) is
decreased by the considered quantum channel on the average, acting on a
given distribution of quantum state or the subsystem of an entangled state.
This treatment has not been strange since the entangling power of a unitary
operator as well as the similar applications in other cases was introduced
[27,28]. However, because the calculation of the abilities of a quantum
channel strongly depends on the structure of the quantum states which
undergo the channel, the direct result is that only 2- dimensional quantum
channel can be effectively considered. For the high dimensional quantum channels, one might have to
consider these behaviors on a concrete state, which is analogous to that the entangling power can be only considered 
for the systems of two qubits [27]. These cases will not be covered in this paper. This paper is organized as
follows. In Sec. II, we treat the quantum channel as the reduction mechanism
of the fidelity and present the infidelity power accompanied by some
concrete examples. In Sec. III, we consider how to influence the coherence
of a state and give the decoherence power. Some examples are also provided.
In Sec. IV, we analyze the potential confusion if we consider the
decoherence of a mixed state and briefly discuss how to consider the
influence of quantum channel on the subsystem of a composite quantum system.
The conclusion is drawn in Sec. V.

\section{Infidelity power}

\subsection{Fidelity}

When a quantum state undergoes a quantum channel, the state will generally
be influenced. Although some particular features of the state could not be
changed, the concrete form of the state, i.e., the fidelity, is usually
changed. In order to give a description of the ability to which degree a
quantum channel influences a quantum state, we would like to first consider
the infidelity power of a quantum channel.

With fidelity of two states $\rho $ and $\rho ^{\prime }$ mentioned, one
could immediately come up with the fidelity defined by $F^{\prime }\left(
\rho ,\rho ^{\prime }\right) =Tr\sqrt{\sqrt{\rho }\rho ^{\prime }\sqrt{\rho }%
}$ or the trace distance defined by $D^{\prime }\left( \rho ,\rho ^{\prime
}\right) =\left\Vert \rho -\rho ^{\prime }\right\Vert _{1}$ with $\left\Vert
A\right\Vert _{1}=Tr\sqrt{AA^{\dagger }}$ [29]. However, consider some given
distribution of state $\rho $, one can find that the mentioned definitions
are not convenient to derive a state-independent quantity. So we would like
to consider another definition of the fidelity based on Frobenius norm $%
\left\Vert \cdot \right\Vert _{2}$.

\textit{Definition 1}.-The fidelity of the state $\rho $ and $\rho ^{\prime
} $ is defined by 
\begin{equation}
F\left( \rho ,\rho ^{\prime }\right) =1-\frac{1}{2}\left\Vert \rho -\rho ^{\prime
}\right\Vert _{2}^{2}.
\end{equation}%
It is clear that if and only if $\rho $ and $\rho ^{\prime }$ are the same,
the fidelity $F=1$. To proceed, we have to introduce a lemma.

\textit{Lemma 1}. For any $n$-dimensional matrix $A$, and an $\left(
n\otimes n\right) $-dimensional maximally entangled state in the
computational basis $\left\vert \Phi _{n}\right\rangle =\sum\limits_{k}\frac{%
1}{\sqrt{n}}\left\vert kk\right\rangle $, the following relations hold:%
\begin{equation}
\left( A\otimes \mathbf{1}_{n}\right) \left\vert \Phi _{n}\right\rangle
=\left( \mathbf{1}_{n}\otimes A^{T}\right) \left\vert \Phi _{n}\right\rangle
,
\end{equation}%
and 
\begin{equation}
TrA=n\left\langle \Phi _{n}\right\vert \left( A\otimes \mathbf{1}_{n}\right)
\left\vert \Phi _{n}\right\rangle .
\end{equation}%
\textit{Proof. }The proof is direct, which is also implied in Ref.
[27].\hfill$\blacksquare$

Now, let $\$$ denote a quantum channel and $\rho ^{\prime }=\$\left( \rho
\right) $ denote the final state of $\rho $ going through the channel, the
fidelity given in Eq. (1) can be rewritten as%
\begin{eqnarray}
F\left( \rho ,\$\left( \rho \right) \right) &=&1-\frac{1}{2}\left\Vert \rho -\$\left(
\rho \right) \right\Vert _{2}^{2}  \notag \\
&=&1-\frac{1}{2}Tr\rho ^{2}-\frac{1}{2}Tr\$\left( \rho \right) ^{2}+Tr\rho \$\left( \rho \right) .
\end{eqnarray}%
Based on Lemma 1, we can find that%
\begin{gather}
Tr\rho \$\left( \rho \right) =n\left\langle \Phi _{n}\right\vert \rho
\$\left( \rho \right) \otimes \mathbf{1}_{n}\left\vert \Phi _{n}\right\rangle
\notag \\
=n\left\langle \Phi _{n}\right\vert \rho \otimes \left[ \$(\rho )\right]
^{T}\left\vert \Phi _{n}\right\rangle  \notag \\
=nTr\left[ \rho \otimes \rho ^{\ast }\right] \left( \$\otimes \mathbf{1}%
_{n}\right) \left\vert \Phi _{n}\right\rangle \left\langle \Phi
_{n}\right\vert  \notag \\
=n^{3}\left\langle \Phi _{n^{2}}\right\vert \left[ \rho \otimes \rho ^{\ast }%
\right] \left( \$\otimes \mathbf{1}_{n}\right) \left\vert \Phi
_{n}\right\rangle \left\langle \Phi _{n}\right\vert \otimes \mathbf{1}%
_{n^{2}}\left\vert \Phi _{n^{2}}\right\rangle  \notag \\
=n^{3}Tr\left( \$\otimes \mathbf{1}_{n}\right) \left\vert \Phi
_{n}\right\rangle \left\langle \Phi _{n}\right\vert \otimes \rho ^{\ast
}\otimes \rho  \notag \\
\times S_{23}\left( \left\vert \Phi _{n}\right\rangle \left\langle \Phi
_{n}\right\vert \otimes \left\vert \Phi _{n}\right\rangle \left\langle \Phi
_{n}\right\vert \right) S_{23}
\end{gather}%
and 
\begin{gather}
Tr\$\left( \rho \right) ^{2}=n\left\langle \Phi _{n}\right\vert \left[
\$(\rho )\right] ^{2}\otimes \mathbf{1}_{n}\left\vert \Phi _{n}\right\rangle
\notag \\
=nTr\left[ \$(\rho )\otimes \rho ^{\ast }\right] \left( \$\otimes \mathbf{1}%
_{n}\right) \left\vert \Phi _{n}\right\rangle \left\langle \Phi
_{n}\right\vert  \notag \\
=n^{3}\left\langle \Phi _{n^{2}}\right\vert \left[ \$(\rho )\otimes \rho
^{\ast }\right] \left( \$\otimes \mathbf{1}_{n}\right) \left\vert \Phi
_{n}\right\rangle \left\langle \Phi _{n}\right\vert \otimes \mathbf{1}%
_{n^{2}}\left\vert \Phi _{n^{2}}\right\rangle  \notag \\
=n^{3}\left\langle \Phi _{n^{2}}\right\vert \left( \$\otimes \mathbf{1}%
_{n}\right) \left\vert \Phi _{n}\right\rangle \left\langle \Phi
_{n}\right\vert \otimes \left[ \$(\rho )\right] ^{T}\otimes \rho \left\vert
\Phi _{n^{2}}\right\rangle  \notag \\
=n^{3}Tr\left\{ \left( \$\otimes \mathbf{1}_{n}\right) \left\vert \Phi
_{n}\right\rangle \left\langle \Phi _{n}\right\vert \otimes \rho ^{\ast
}\otimes \rho \right\}  \notag \\
\times S_{23}\left( \$\otimes \mathbf{1}_{n}\right) \left\vert \Phi
_{n}\right\rangle \left\langle \Phi _{n}\right\vert \otimes \left\vert \Phi
_{n}\right\rangle \left\langle \Phi _{n}\right\vert S_{23},
\end{gather}%
where we consider $\left\vert \Phi _{n^{2}}\right\rangle =S_{23}\left\vert
\Phi _{n}\right\rangle \left\vert \Phi _{n}\right\rangle $ with $S_{23}$
representing the swapping operations between the \textit{2nd} and \textit{3rd%
} subsystems. Let%
\begin{equation}
\left\{ 
\begin{array}{c}
W_{1}=S_{23}\left[ \tilde{\$}\left( \varrho _{\Phi }\right) \otimes \varrho
_{\Phi }\right] S_{23} \\ 
W_{2}=S_{23}\left[ \varrho _{\Phi }\otimes \varrho _{\Phi }\right] S_{23} \\ 
Q_{1}=\tilde{\$}\left( \varrho _{\Phi }\right) \otimes \mathbf{1}_{n^{2}} \\ 
Q_{2}=\varrho _{\Phi }\otimes \mathbf{1}_{n^{2}}%
\end{array}%
\right.
\end{equation}%
with $\tilde{\$}\left( \varrho _{\Phi }\right) =\left( \$\otimes \mathbf{1}%
_{n}\right) \left\vert \Phi _{n}\right\rangle \left\langle \Phi
_{n}\right\vert $ and $\varrho _{\Phi }=\left\vert \Phi _{n}\right\rangle
\left\langle \Phi _{n}\right\vert $, and substitute Eq. (5) and Eq. (6) into
Eq. (4), $F\left( \rho ,\$\left( \rho \right) \right) $ can be rewritten as%
\begin{equation}
F\left( \rho ,\$\left( \rho \right) \right) =1-\frac{n^{3}}{2}Tr\left( \mathbf{1}%
_{n^{2}}\otimes \rho ^{\ast }\otimes \rho \right) \mathcal{M}.
\end{equation}%
with 
\begin{equation}
\mathcal{M}=\left[ Q_{1}(W_{1}-W_{2})+\left( Q_{2}-Q_{1}\right) W_{2}\right]
.
\end{equation}%
Thus $\mathcal{M}$ only depends on the influence of the maximally entangled
state $\left\vert \Phi _{n}\right\rangle $ through the considered quantum
channel instead of the information of the state of interests. The infidelity
power can be directly defined as the difference of fidelity before and after
going through the channel subject to some given distribution. However,
different distributions would lead to quite different infidelity power. In
this paper, we restrict ourselves to the uniform distribution.

\subsection{Infidelity power}

In order to describe the difference between the fidelities of the state $%
\rho $ with and without quantum channel $\$$, we can use 
\begin{eqnarray}
\tilde{e}_{F}\left( \rho ,\$\left( \rho \right) \right) &=&F\left( \rho
,\rho \right) -F\left( \rho ,\$\left( \rho \right) \right)  \notag \\
&=&\frac{n^{3}}{2}Tr\left( \mathbf{1}_{n^{2}}\otimes \rho ^{\ast }\otimes \rho \right) 
\mathcal{M}.
\end{eqnarray}%
Consider the uniform distribution $\mathcal{F}$ of the state $\rho $ [27],
we can define the infidelity power as follows.

\textit{Definition 2}. The infidelity power of the $n$-dimensional quantum
channel $\$$ can be defined as%
\begin{equation}
e_{\mathcal{F}}\left( \$\right) =\frac{n^{3}}{2}Tr\left( \mathbf{1}_{n^{2}}\otimes 
\mathcal{F}(\rho )\right) \mathcal{M},
\end{equation}%
with $\mathcal{F}(\rho )=\frac{1}{\Omega _{\mathcal{F}}}\int_{\mathcal{F}%
}\rho ^{\ast }\otimes \rho du(\rho )$ where $\Omega _{\mathcal{F}}$ is some
normalization factor, $du(\rho )$ denotes the measure over the state $\rho $
induced by the uniform distribution $\mathcal{F}$.

It is obvious that $\mathcal{F}(\rho )$ does not depend on the concrete form
of a density matrix but only the dimension of the density matrices and the
structure of the states. However, it is unfortunate that the structure of
the high dimensional states is very complicated. It is difficult to describe
a given distribution of such a state space. Therefore, we only present the
concrete expression of $\mathcal{F}(\rho )$ for a system of qubit.

\textit{Theorem 1}. The infidelity power of the quantum channel $\$$ on a
qubit can be given by%
\begin{equation}
\tilde{e}_{\mathcal{F}}\left( \$\right) =4Tr\left( \mathbf{1}_{n^{2}}\otimes 
\mathcal{F}(\rho )\right) \mathcal{M},
\end{equation}%
with 
\begin{equation}
\mathcal{F}(\rho )=\frac{1}{5}\mathbf{1}_{4}\otimes \left( \mathbf{1}%
_{4}+\left\vert \Phi _{2}\right\rangle \left\langle \Phi _{2}\right\vert
\right) .
\end{equation}

\textit{Proof.} The density matrix of any qubit $\rho $ can be given in the
Bloch representation as%
\begin{equation}
\rho =\frac{1}{2}\left( \mathbf{1}_{2}+r\mathbf{n}\cdot \mathbf{\sigma }%
\right)
\end{equation}%
where $\mathbf{\sigma }=[\sigma _{x},\sigma _{y},\sigma _{z}]$ is the Pauli
matrices, $\mathbf{n}=[\sin \theta \cos \varphi ,\sin \theta \sin \varphi
,\cos \theta ]$ with $\theta $ the inclination and $\varphi $ the azimuth in
the Bloch sphere, respectively, and $r$ is the radius. Since we suppose the
state is distributed uniformly, the state density can be given by $r^{2}\sin
\theta $. So one can easily find that%
\begin{equation}
\Omega _{\mathcal{F}}=\int\limits_{0}^{1}r^{2}dr\int_{0}^{\pi }\sin \theta
d\theta \int_{0}^{2\pi }d\varphi =\frac{4\pi }{3},
\end{equation}%
and 
\begin{eqnarray}
\mathcal{F}(\rho ) &=&\frac{3}{16\pi }\int\limits_{0}^{1}r^{2}dr\int_{0}^{%
\pi }\sin \theta \int_{0}^{2\pi }\left[ \mathbf{1}_{4}\right. +r\mathbf{n}%
\cdot \left( \mathbf{1}_{2}\otimes \mathbf{\sigma }+\mathbf{\sigma }\otimes 
\mathbf{1}_{2}\right)  \notag \\
&&+r^{2}\left. \left( \mathbf{n}\otimes \mathbf{n}\right) \cdot \left( 
\mathbf{\sigma }^{\ast }\otimes \mathbf{\sigma }\right) \right] d\varphi
d\theta  \notag \\
&=&\frac{1}{4}\left[ \mathbf{1}_{4}+\frac{1}{5}\left( \sigma _{x}\otimes
\sigma _{x}-\sigma _{y}\otimes \sigma _{y}+\sigma _{z}\otimes \sigma
_{z}\right) \right] ,
\end{eqnarray}%
which is consistent with Eq. (13). The proof is completed.\hfill$\blacksquare
$

\subsection{Examples}

As applications, we would like to calculate the infidelity power for a qubit
of the depolarizing channel $\$_{d}$ given in Kraus representation as 
\begin{equation}
M_{0}^{d}=\sqrt{1-p}\mathbf{1}_{2},M_{k}^{d}=\sqrt{\frac{p}{3}}\sigma
_{k},k=1,2,3,
\end{equation}%
the phase-damping channel $\$_{p}$ given by 
\begin{eqnarray}
M_{0}^{p} &=&\sqrt{1-p}\mathbf{1}_{2}, \\
M_{1}^{p} &=&\sqrt{p}\left( 
\begin{array}{cc}
1 & 0 \\ 
0 & 0%
\end{array}%
\right) , \\
M_{2}^{p} &=&\sqrt{p}\left( 
\begin{array}{cc}
0 & 0 \\ 
0 & 1%
\end{array}%
\right) ,
\end{eqnarray}%
the amplitude-damping channel $\$_{a}$ given by 
\begin{eqnarray}
M_{0}^{a} &=&\left( 
\begin{array}{cc}
1 & 0 \\ 
0 & \sqrt{1-p}%
\end{array}%
\right) , \\
M_{1}^{a} &=&\left( 
\begin{array}{cc}
0 & \sqrt{p} \\ 
0 & 0%
\end{array}%
\right) ,
\end{eqnarray}%
and the generalized amplitude-damping channel $\$_{g}$ given by%
\begin{eqnarray*}
M_{0}^{g} &=&\sqrt{q}M_{0}^{a},M_{1}^{g}=\sqrt{q}M_{1}^{a}, \\
M_{2}^{g} &=&\sqrt{1-q}\left( M_{1}^{a}\right) ^{\dag },M_{3}^{g}=\sqrt{1-q}%
\sigma _{x}M_{0}^{a}\sigma _{x}.
\end{eqnarray*}%
Substitute these quantum channels [29] into $\$_{i}\left( \rho _{\Phi
}\right) $, one can easily find that 
\begin{equation}
\tilde{\$}_{d}\left( \varrho _{\Phi }\right) =\left[ 
\begin{array}{cccc}
\frac{1}{2}-\frac{p}{3} & 0 & 0 & \frac{1}{2}-\frac{2p}{3} \\ 
0 & \frac{p}{3} & 0 & 0 \\ 
0 & 0 & \frac{p}{3} & 0 \\ 
\frac{1}{2}-\frac{2p}{3} & 0 & 0 & \frac{1}{2}-\frac{p}{3}%
\end{array}%
\right] ,
\end{equation}%
\begin{equation}
\tilde{\$}_{p}\left( \varrho _{\Phi }\right) =\left[ 
\begin{array}{cccc}
\frac{1}{2} & 0 & 0 & \frac{1-p}{2} \\ 
0 & 0 & 0 & 0 \\ 
0 & 0 & 0 & 0 \\ 
\frac{1-p}{2} & 0 & 0 & \frac{1}{2}%
\end{array}%
\right] ,
\end{equation}%
\begin{equation}
\tilde{\$}_{a}\left( \varrho _{\Phi }\right) =\left[ 
\begin{array}{cccc}
\frac{1}{2} & 0 & 0 & \frac{\sqrt{1-p}}{2} \\ 
0 & \frac{p}{2} & 0 & 0 \\ 
0 & 0 & 0 & 0 \\ 
\frac{\sqrt{1-p}}{2} & 0 & 0 & \frac{1-p}{2}%
\end{array}%
\right] ,
\end{equation}%
and%
\begin{equation}
\tilde{\$}_{g}\left( \varrho _{\Phi }\right) =\left[ 
\begin{array}{cccc}
\frac{1-p+pq}{2} & 0 & 0 & \frac{\sqrt{1-p}}{2} \\ 
0 & \frac{pq}{2} & 0 & 0 \\ 
0 & 0 & \frac{p(1-q)}{2} & 0 \\ 
\frac{\sqrt{1-p}}{2} & 0 & 0 & \frac{1-pq}{2}%
\end{array}%
\right] ,
\end{equation}%
Consider the swap operator $S_{23}=\mathbf{1}_{2}\otimes
\sum\limits_{j,k=0}^{1}\left\vert jk\right\rangle \left\langle kj\right\vert
\otimes $ $\mathbf{1}_{2}$ and substitute Eqs (23-26) into Eq. (12), we have

\begin{eqnarray}
\tilde{e}_{\mathcal{F}}\left( \$_{d}\right) &=&\frac{4p^{2}}{15}, \\
\tilde{e}_{\mathcal{F}}\left( \$_{p}\right) &=&\frac{p^{2}}{10}, \\
\tilde{e}_{\mathcal{F}}\left( \$_{a}\right) &=&\frac{p(3p-1)}{10}+\frac{(1-%
\sqrt{1-p})}{5}, \\
\tilde{e}_{\mathcal{F}}\left( \$_{g}\right) &=&\tilde{e}_{\mathcal{F}}\left(
\$_{a}\right) -p^{2}q(1-q).
\end{eqnarray}%
It is apparent that $\tilde{e}_{\mathcal{F}}\left( \$_{i}\right) =0$ for $%
p=0 $, which means there is no quantum channel operating on them. $\tilde{e}%
_{\mathcal{F}}\left( \$_{g}\right) =\tilde{e}_{\mathcal{F}}\left(
\$_{a}\right) $, if $q=0$, or $1$. This is because $\$_{g}$ will become $%
\$_{a}$ or its dual quantum channel on this condition. It is a natural
conclusion that $\$_{i}$ reduce the fidelity due to $\tilde{e}_{\mathcal{F}%
}\left( \$_{i}\right) >0$, in particular, with the increase of the decay
rate $p$, the infidelity power is increasing. But it is not difficult to
find that given $p$, the ability of reducing the fidelity can be weakened if
we adjust $q$ for the generalized amplitude-damping channel. The minimal
infidelity power is obtained if $q=\frac{1}{2}$.

\section{Decoherence power}

\subsection{Coherence}

For a given density matrix, the definition of quantum coherence depends on
not only the density matrix itself, but also the bases on which the density
matrix is written. To evaluate whether there exists quantum coherence in a
quantum state, physically one has to find some observables to reveal the
interference by measuring them. It can be shown that all the quantum states
(density matrices) but the maximally mixed state can demonstrate
interference because one can always find such an observable that can reveal
it so long as the observable doesn't commute with the density matrix of
interests. Based on such a think, one could find various ways to defining
quantum coherence. Here we would like to define the maximal distance between
the density matrix of interests and the corresponding maximally mixed state
with all potential bases taken into account. The rigorous formulae can be
given as follows.

\textit{Definition 3}.-Quantum coherence for an $n$-dimensional density
matrix $\rho $ is defined as the maximal distance between $\rho $ and the
maximally mixed state $\frac{\mathbf{1}_{n}}{n}$ with all potential bases
considered, i.e.,

\begin{equation}
D(\rho )=Tr\rho ^{2}-\frac{1}{n},
\end{equation}%
where $\mathbf{1}_{n}$ denotes the $n$-dimensional identity. $D(\rho )$ is
also the maximally potential coherence within all possible bases.

\textit{Proof. }The density matrix $\rho $ in a different framework can be
given by $U\rho U^{\dagger }$ where $U$ is the unitary transformation
relating two different bases of $\rho $. So the distance $D(\rho )$ can be
written as%
\begin{eqnarray}
D(\rho ) &=&\max_{U}\left\Vert U\rho U^{\dagger }-\frac{1}{n}\right\Vert
_{2}^{2}  \notag \\
&=&\max_{U}Tr\left( U\rho U^{\dagger }-\frac{1}{n}\right) ^{2}  \notag \\
&=&Tr\rho ^{2}-\frac{1}{n},
\end{eqnarray}%
where $\left\Vert \cdot \right\Vert $ is the Frobenius norm.

In other words, given a basis in which the density matrix is written by $%
\rho $, the coherence in this basis can be described completely by the
off-diagonal entries of $\rho$ [11]. So if we extract the maximal
contribution of coherence with different basis considered, we have 
\begin{eqnarray}
D(\rho ) &=&\max_{U}\sum_{i\neq j}\left[ U\rho U^{\dagger }\right] _{ij}^{2}
\notag \\
&=&Tr\rho ^{2}-\min_{U}\sum_{k}\left[ U\rho U^{\dagger }\right] _{kk}^{2} 
\notag \\
&=&Tr\rho ^{2}-\frac{1}{n}.
\end{eqnarray}%
Eq. (33) holds because one can always find a unitary matrix such that $U\rho
U^{\dagger }$ have the equal diagonal entries. It is obviously shown that
our definition actually quantify the maximal coherence of a state by
considering all the possible bases [11]. The proof is completed.\hfill$%
\blacksquare$

From the definition, one will easily see that such a distance does not
depend on the bases, i.e., the unitary matrix $U$. This does not contradict
with the usual statement of the dependence of bases for quantum coherence.
In fact, the dependence of bases is in that the different observables chosen
to reveal the interference will lead to the different interference
visibilities. This actually implies the ability of the observable that can
reveal the quantum coherence of the given state, instead how much quantum
coherence could be revealed for a state. In addition, it is obvious that the
maximal value of the coherence measure is $\frac{n-1}{n}$ which can be
attained by all the pure states. It is easy to understand because all the
pure states are equivalent or interconverted under appropriate unitary
transformations. It is obvious that Eq. (31) is also closely related to the
purity of a quantum state, the quantumness of a single state, so the
coherence measure can also be understood as the purity measure or the
quantumness measure etc. with a small potential deformation [30,31].

With this definition, we can proceed to consider a quantum system with the
state $\rho $ undergoes a quantum channel $\$$ with the final state can be
given by $\$(\rho )$. Thus the coherence of the final state $\$(\rho )$ can
be easily written as%
\begin{equation}
D(\$(\rho ))=Tr\left[ \$(\rho )\right] ^{2}-\frac{1}{n}.
\end{equation}%
For the latter use, next we would like to present a new form that separates
the quantum state and the quantum channel in Eq. (34).

Using Eqs. (6) and (7), one can directly obtain%
\begin{equation}
D(\$(\rho ))=n^{3}Tr\left( \mathbf{1}_{n^{2}}\otimes \rho ^{\ast }\otimes
\rho \right) Q_{1}W_{1}-\frac{1}{n}.
\end{equation}%
Thus Eq. (35) shows that the left quantum coherence of $\rho $ when it
passes through the quantum channel $\$$. So the decoherence power that
describes to what degree the decoherence has been reduced can be defined as
the difference of the coherence before and after the channel subject to some
distribution of quantum states.

\subsection{Definition of decoherence power}

Based on the above analysis, we can obtain the following definition for
decoherence power.

\textit{Definition 4}. The decoherence power of the $n$-dimensional quantum
channel $\$$ can be defined as%
\begin{equation}
e_{\mathcal{D}}\left( \$\right) =1-n^{3}Tr\left( \mathbf{1}_{n^{2}}\otimes 
\mathcal{D}(\rho )\right) Q_{1}W_{1},
\end{equation}%
with $\mathcal{D}(\rho )=\frac{1}{\Omega _{\mathcal{D}}}\int_{\mathcal{D}%
}\rho ^{\ast }\otimes \rho du(\rho )$ where $\Omega _{\mathcal{D}}$ $du(\rho
)$ denotes the measure over the pure state $\rho $ induced by the uniform
distribution $\mathcal{D}$.

From this definition, we should first note that we only consider the
distribution of pure states instead of the whole state space. Why we don't
cover mixed states will be analyzed in the latter part. In addition, one can
find that $\mathcal{D}(\rho )$ analogous to infidelity power does not depend
on the concrete form of a density matrix but only the dimension of the
density matrices and its structure. However, due to the same reason as the
infidelity power, we only present the concrete expression of $\mathcal{D}%
(\rho )$ for a system of qubit.

\textit{Theorem 2}. The decoherence power of the quantum channel $\$$ on a
qubit can be given by%
\begin{equation}
\tilde{e}_{\mathcal{D}}\left( \$\right) =1-8TrDQ_{1}W_{1},
\end{equation}%
with 
\begin{equation}
\mathcal{D}(\rho )=\frac{1}{3}\mathbf{1}_{4}\otimes \left( \frac{\mathbf{1}%
_{4}}{2}+\left\vert \Phi _{2}\right\rangle \left\langle \Phi _{2}\right\vert
\right) .
\end{equation}

\textit{Proof.} The density matrix of any pure qubit can be given in the
Bloch representation as%
\begin{equation}
\rho =\frac{1}{2}\left( \mathbf{1}_{2}+\mathbf{n}\cdot \mathbf{\sigma }%
\right)
\end{equation}%
where $\mathbf{\sigma }=[\sigma _{x},\sigma _{y},\sigma _{z}]$ is the Pauli
matrices, $\mathbf{n}=[\sin \theta \cos \varphi ,\sin \theta \sin \varphi
,\cos \theta ]$ with $\theta $ the inclination and $\varphi $ the azimuth in
the Bloch sphere, respectively. Since we suppose the state is distributed
uniformly, the state density can be given by $\sin \theta $. So one can
easily find that%
\begin{equation}
\Omega =\int_{0}^{\pi }\sin \theta d\theta \int_{0}^{2\pi }d\varphi =4\pi .
\end{equation}%
Thus, 
\begin{equation}
\mathcal{D}(\rho )=\frac{1}{4}\mathbf{1}_{4}+\frac{1}{12}\left( \sigma
_{x}\otimes \sigma _{x}-\sigma _{y}\otimes \sigma _{y}+\sigma _{z}\otimes
\sigma _{z}\right) .
\end{equation}%
It is equivalent to Eq. (38). The proof is completed.\hfill$\blacksquare$

\subsection{Examples}

Similarly, we also calculate the decoherence power for a qubit of the
depolarizing channel $\$_{d}$, the phase-damping channel $\$_{p}$, the
amplitude-damping channel $\$_{a}$ and the generalized amplitude-damping
channel $\$_{g}$. Based on the analogous calculation to those in Eqs.
(27-30), we have 
\begin{eqnarray}
\tilde{e}_{\mathcal{D}}\left( \$_{d}\right) &=&\frac{4p}{3}\left( 1-\frac{2p%
}{3}\right) , \\
\tilde{e}_{\mathcal{D}}\left( \$_{p}\right) &=&\frac{p}{3}\left( 2-p\right) ,
\\
\tilde{e}_{\mathcal{D}}\left( \$_{a}\right) &=&\frac{2p}{3}(1-p), \\
\tilde{e}_{\mathcal{D}}\left( \$_{g}\right) &=&\tilde{e}_{\mathcal{D}}\left(
\$_{a}\right) +2p^{2}q(1-q).
\end{eqnarray}%
As is expected again, we can find that $\tilde{e}_{\mathcal{D}}\left(
\$_{i}\right) $ will vanish if $p=0$. However, unlike the infidelity power
which is a monotone function of $p$, the decoherence power of all the
mentioned channels but the phase-damping channel will reach a maximum at
some particular $p^{\ast }\in \left[ 0,1\right] $. This should be
distinguished from the case where only the coherence in a fixed basis is
considered.

\section{Decoherence of mixed states}

\subsection{Confusion of decoherence power based on mixed states}

In fact, Eq. (35) is suitable for all quantum states including mixed states
when we consider the decoherence of a state through a channel. However, it
is not hard to see that in the sense of the previous definition of
coherence, the reduction of coherence depends on not only the quantum
channel itself, but also the quantum state that undergoes the channel. So it
is possible that a given quantum channel could reduce the coherence of some
states and increase the coherence of the other states. Thus the decoherence
power will lead to confusion since it is defined as the average contribution
of the reduction of coherence subject to a given state distribution. In
order to demonstrate the inconsistent roles of a quantum channel on
different states, we would like to take the above three quantum channels as
examples to show the reduction and increase of the coherence.

Actually, one can find that the coherence of any states based on definition
1 will be reduced if they undergo the depolarizing channel $\$_{d}$ and the
phase-damping channel $\$_{p}$. But the amplitude-damping channel could lead
to the increase of the coherence. In Fig. 1, we plot the two regions within
and without which the coherence of all the states can be reduced and
increased, respectively.

\begin{figure}[tbp]
\centering
\includegraphics[width=0.6\columnwidth]{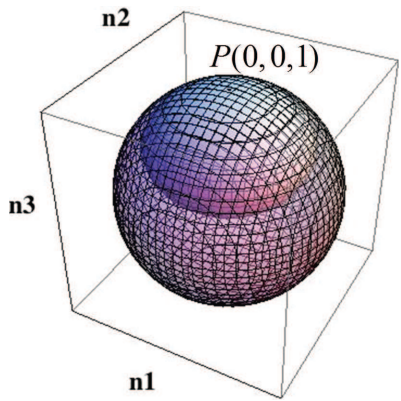}
\caption{Qubits in Bloch representation undergoing the amplitude-damping
channel. For the channel, $p=0.1$. The outer ball represents all the states
of a qubit. The inner spheriod denotes the states that loses coherence via
the channel. The two balls inscribe at point $P=(0,0,1)$ }
\end{figure}

\subsection{Quantum channels on subsystems: disentangling factor}

Since the direct consideration of decoherence of mixed state could lead to
the confusion, we have to consider the mixed states in an indirect way,
namely, we turn to an entangled bipartite composite system of a pure state
with one of its subsystems undergoing a quantum channel. It is obvious that
the reduced density matrix of any subsystem is mixed; In addition, any local
quantum channel on the subsystem will always reduce the entanglement of the
composite system since a good entanglement measure should be an entanglement
monotone. However, it should be noted that the decoherence based on
disentanglement is different from our initial definition 1. Strictly
speaking, we actually characterize the reduction of a special
coherence------quantum entanglement. In this sense, what we consider should
be called as the disentangling power instead of the decoherence power of the
original definition 1.

Consider a bipartite pure state of two qubits $\rho _{AB}$, the entanglement
can be reduced if one subsystem undergoes a quantum channel $\$$. If we
employ concurrence [32] as entanglement measure, one can easily find that
[33] 
\begin{equation*}
C(\$\left( \rho _{AB}\right) )=C(\$\left( \rho _{\Phi }\right) )C(\rho
_{AB}),
\end{equation*}%
with $\rho _{\Phi }=\left\vert \Phi _{2}\right\rangle \left\langle \Phi
_{2}\right\vert $. So no matter what distribution of quantum states is
considered, the concurrence is directly reduced by the factor $C(\$\left(
\rho _{\Phi }\right) )$. Thus the disentangling power of $\$$ in this case
can be directly characterized by $C(\$\left( \rho _{\Phi }\right) )$.

In addition, if $\rho_{AB}$ is a mixed state, one can also consider the
reduction of entanglement subject to some quantum channel. However, due to
the complicated expression of concurrence for mixed states. So far there has
not such a factorial form of the reduction of entanglement.

\section{Conclusions}

In this paper, we study how a quantum channel influences the fidelity and
the coherence of a state when the state goes through it and briefly discuss
the reduction of entanglement when a subsystem undergoes a channel. We give
the infidelity power and decoherence power of a quantum channel. They are
independent of quantum states and describe an average contribution of the
infidelity and the decoherence. As applications, we calculate the infidelity
power and decoherence power of depolarizing channel, phase-damping channel,
amplitude-damping channel and generalized amplitude-damping channel,
respectively. We show that although quantum channels (if it is not trivial)
definitely reduce the fidelity of the state through it and the entanglement
with subsystems through it, for some channels, they could increase the
coherence of some states.

\section{Acknowledgements}

This work was supported by the National Natural Science Foundation of China,
under Grant No. 11175033 and the Fundamental Research Funds of the Central
Universities, under Grant No. DUT12LK42.

\end{document}